\documentstyle[preprint,aps]{revtex}

\tightenlines

\begin{document}
\title{Disorder and thermally driven vortex-lattice melting in La$_{2-x}$Sr$%
_{x}$CuO$_{4}$ crystals}
\author{Y. Radzyner, A. Shaulov, Y. Yeshurun}
\address{Department of Physics, Institute of Superconductivity,\\
Bar-Ilan University, Ramat-Gan, Israel}
\author{I. Felner}
\address{Raccah Institute of Physics, The Hebrew University,\\
Jerusalem 91904, Israel }
\author{K. Kishio and J. Shimoyama}
\address{Department of Applied Chemistry, The University of Tokyo,\\
Tokyo 113-8656, Japan }
\date{\today}
\maketitle

\begin{abstract}
Magnetization measurements in La$_{2-x}$Sr$_{x}$CuO$_{4}$ crystals indicate
vortex order-disorder transition manifested by a sharp kink in the second
magnetization peak. The transition field exhibits unique temperature
dependence, namely a strong decrease with temperature in the entire measured
range. This behavior rules out the conventional interpretation of a
disorder-driven transition into an entangled vortex solid phase. It is shown
that the transition in La$_{2-x}$Sr$_{x}$CuO$_{4}$ is driven by {\it both}
thermally- and disorder-induced fluctuations, resulting in a pinned liquid
state. We conclude that vortex solid-liquid, solid-solid and solid to
pinned-liquid transitions are different manifestations of the same
thermodynamic order-disorder transition, distinguished by the relative
contributions of thermal and disorder-induced fluctuations.
\end{abstract}

\pacs{74.60.Ge, 74.72.Dn}

The nature of the various vortex matter phases in high-temperature
superconductors (HTS), and the transitions between them, have been the topic
of many experimental and theoretical investigations \cite%
{zeldov1,khaykovich96,giller97,deligiannis,giller99,sasagawa,giamarchi,ertas,vinokur,blatter}%
. Two vortex order-disorder phase transitions have been identified: A
melting transition into a liquid vortex state manifested by a discontinuous
jump in the reversible magnetization \cite{zeldov1}, and a solid-solid
transition into an entangled vortex state \cite{khaykovich96,giller97}
manifested by the appearance of a second magnetization peak with pronounced
features (onset \cite{khaykovich96,giller97}, kink \cite%
{giller99,nishizaki2000} or peak \cite{deligiannis}). Theoretical treatments
attempting to describe the vortex phase diagram in HTS \cite%
{giamarchi,ertas,vinokur,blatter}, ascribe the melting transition to thermal
fluctuations and the solid-solid transition to disorder induced fluctuations
of vortices. Accordingly, the melting line is determined by the competition
between the elastic energy, $E_{el}$, and the thermal energy, $kT$, while
the contest between $E_{el}$ and the pinning energy, $E_{pin}$, determines
the solid-solid transition line. The melting line is expected to decrease
strongly with temperature as thermal fluctuations are enhanced, whereas the
vortex solid-solid transition line is expected to maintain a constant value
at low temperatures where both $E_{pin}$ and $E_{el}$ become temperature
independent. Experiments in a variety of HTS crystals \cite%
{khaykovich96,giller97,giller99,sasagawa,baziljevich} basically conform to
this theory, yielding a melting line which decreases with temperature, or a
vortex solid-solid transition line which is temperature independent in a
wide range of temperatures.

In this paper we report on a significantly different behavior obtained in La$%
_{2-x}$Sr$_{x}$CuO$_{4}$ (LaSCO) crystals. Magnetization measurements reveal
a transition of a quasi-ordered vortex lattice into a disordered vortex
state with enhanced vortex pinning, indicated by a sharp kink in the second
magnetization peak \cite{giller99,nishizaki2000}. However, the transition
field exhibits a unique behavior, namely strong temperature dependence in
the entire measured range. This behavior rules out the conventional
interpretation of a transition into an entangled solid vortex phase in which
only $E_{pin}$ and $E_{el}$ play a role. We demonstrate that in order to
explain the behavior of the transition line in LaSCO, one must take into
account the contribution of thermal energy as well. Thus, our LaSCO samples
provide a unique example where the transition to the vortex disordered state
is driven by {\it both} thermally- and disorder-induced fluctuations. The
resulting disordered state may be identified as a liquid state with
irreversible magnetic behavior, i.e. a vortex pinned-liquid state \cite%
{footnote3}.

Several samples were cut from a single (La$_{0.937}$Sr$_{0.063}$)$_{2}$CuO$%
_{4}$ crystal, with T$_{c}$ of about 32 K. Data will be shown for sample L1 (%
$0.8\times 2.5\times 0.8$ $mm^{3}$), though all samples give similar results
in all aspects. Magnetization measurements were performed using a commercial
SQUID magnetometer (Quantum Design MPMS-5S).

The inset to Fig. 1 presents magnetization loops measured at several
temperatures, with the field parallel to the {\it ab} planes. Similarly to
untwinned YBa$_{2}$Cu$_{3}$O$_{7-\delta }$ (YBCO) \cite{radzyner}, one
observes four distinct features (indicated by arrows): The onset of a second
peak on the ascending branch at $H_{onset}^{+}$, a sharp change in slope of
the magnetization at $H_{kink}^{+}$ \cite{togawa}, and their counterparts on
the descending branch at $H_{onset}^{-}$\ and $H_{kink}^{-}$, respectively.
The temperature dependence of these features is depicted in the main panel
of Fig. 1. Note that all four lines show similar behavior, namely a steep 
{\it concave} descent with the increase of temperature. Similar strong
temperature dependence of $H_{onset}^{+}$ and $H_{onset}^{-}$, was observed
also for $H||c$. However, for $H||c$, $H_{kink}^{+}$ and $H_{kink}^{-}$ were
more difficult to resolve due to the presence of twin boundaries \cite%
{birgeneau}. In this manuscript we therefore focus on results obtained with $%
H||ab$ \cite{footnote1}.

Magnetic relaxation measurements yield further insight into the nature of
these lines. Figure 2 depicts the evolution of the magnetization at 12 K. In
this figure every column represents measurement extended over an hour; the
solid lines in the figure connect values obtained at $t=0$ and $t=1$ hr.
Positions of both $H_{kink}^{+}$ and $H_{kink}^{-}$\ (not shown) do not vary
with time, while both $H_{onset}^{+}$ and $H_{onset}^{-}$\ (not shown)
decrease appreciably over an hour. These observations point to either of the
kink fields, rather than the onset, as indicating an order-disorder
transition, as previously found in YBCO \cite{giller99}. This result is
further refined by measurements of the field dependence of the normalized
magnetic relaxation rate, $s=d(\ln m)/d(\ln t)$ , as depicted in the inset
to Fig. 2: A sharp change in the slope of $s${\it \ vs.} field is observed
at a field corresponding to $H_{kink}^{-}$, on both the branches \cite%
{footnote2}.

The magnetization curves and relaxation data indicate an order-disorder
phase transition of the vortex system occurring at $H_{kink}^{-}$, in
agreement with observations in YBCO \cite{radzyner}. Since the disordered
phase is magnetically irreversible, it is tempting to identify this
transition as a vortex solid-solid phase transition, similar to that
observed in YBCO \cite{giller99}, Bi$_{2}$Sr$_{2}$CaCu$_{2}$O$_{8}$ (BSCCO) %
\cite{khaykovich96}, Nd$_{1.85}$Ce$_{0.15}$CuO$_{4-\delta }$ \cite{giller97}%
\ and Bi$_{1.6}$Pb$_{0.4}$Sr$_{2}$CaCu$_{2}$O$_{8+\delta }$ \cite%
{baziljevich}. We note, however, that contrary to these materials, which
exhibit a temperature independent solid-solid transition line for a wide
range of temperatures, in LaSCO, this line is strongly temperature dependent
in the entire measured temperature range. Thus, the conventional
interpretation of a disordered-driven transition into an entangled solid
phase is refutable.

The measured temperature dependence of the transition line may be influenced
by effects of surface barriers, which might obscure the features of the
second peak anomaly at low temperatures \cite%
{giller97,baziljevich,radzyner,deandrade}. Indeed, magnetization loops in
LaSCO reveal a strong temperature dependence of the field where flux
initially penetrates the sample overcoming surface barriers \cite{deandrade}%
. However, Bean-Livingston barriers play a role only in the increasing
branch of the loop \cite{bean}, and have no effect on the decreasing branch;
the fact that in LaSCO the strong temperature dependence is common to the
features measured on both ascending and descending branches, excludes an
explanation associated with surface barriers.

Another possible explanation for the behavior of the measured transition
line in LaSCO may be associated with the influence of the persistent
current: Strong currents may have a tendency to order the vortices \cite%
{koshelev}, so that transition into a vortex glass would be deferred to
higher fields. As temperature is decreased current increases, and its
influence on the transition line should be marked. This explanation is
precluded by the fact that the position of the kink is unaffected by the
change in current; As can be seen from Fig. 2, within the time window of the
measurement, the current relaxes to about 75\% of its initial value, but the
position of the kink is not altered, while within the same time window the
onset field shifts by about 1 kOe.

In the following, we propose an explanation for the unique temperature
dependence of the transition line measured in LaSCO asserting that this
transition is driven by {\it both} thermally- and disorder-induced
fluctuations. The transition field at $H_{kink}^{-}(T)$ is associated with
the second magnetization peak, as does the solid-solid transition field, but
depends strongly on temperature like the melting field. This strong
temperature dependence implies that the transition to the disordered vortex
state is driven not only by disorder-induced fluctuations, which are
temperature independent far below T$_{c}$, but also by thermal fluctuations.
As both thermal and disorder-induced fluctuations take a part in
destabilizing the ordered solid, the interplay between all three energy
scales $E_{el}$, $E_{pin}$ and $kT$, should determine the transition line %
\cite{vinokurUN,goldscmidt,kierfeld}. The basic premise of our analysis is
that an order-disorder transition occurs when the sum of the thermal and the
disorder-induced displacements of the flux line, 
\mbox{$<$}%
u$_{T}^{2}$%
\mbox{$>$}
and 
\mbox{$<$}%
u$_{dis}^{2}$%
\mbox{$>$}%
, respectively, exceeds a certain fraction of the vortex lattice constant %
\cite{giamarchi}, $a_{o}$. This leads to $%
<u_{T}^{2}>+<u_{dis}^{2}>=c_{L}^{2}a_{o}^{2}$\ ($c_{L}$ is the Lindemann
number), or equivalently \cite{radzyner2} to the energy balance at the
transition field:

\qquad \qquad \qquad \qquad (1) $\ \ E_{el}=E_{pin}+kT$.\ 

More accurate analysis should involve the averaged total displacement of the
flux line, which is not necessarily the sum of 
\mbox{$<$}%
u$_{T}^{2}>$\ and 
\mbox{$<$}%
u$_{dis}^{2}>$. Our simplified analysis yields, however, a qualitative
description, and provides important insight.

We\ numerically solve Eq. (1), using $E_{el}=\varepsilon \epsilon
_{o}c_{L}^{2}a_{o}$\ and $E_{pin}=U_{dp}\left( L_{o}/L_{c}\right) ^{1/5}$\
from the cage model\ \cite{ertas,vinokur}. Here, $\varepsilon $ is the
anisotropy ratio, $\epsilon _{o}=\left( \Phi _{o}/4\pi \lambda \right) ^{2}$%
\ is the vortex line tension, $U_{dp}=\left( \gamma \varepsilon ^{2}\epsilon
_{o}\xi ^{4}\right) ^{1/3}$ is\ the single vortex depinning energy, $%
L_{o}=2\varepsilon a_{o}$ is the characteristic length for the longitudinal
fluctuations, and $L_{c}=\left( \varepsilon ^{4}\epsilon _{o}^{2}\xi
^{2}/\gamma \right) ^{1/3}$\ is the size of the coherently pinned segment of
the vortex. The above expressions for $E_{el}$\ and $E_{pin}$\ are clearly
applicable for analyzing our results for $H||c$. We adopt the same
expressions also for $H||ab$, assuming that Abrikosov vortices, rather than
Josephson vortices, are involved, owing to the small value of the
anisotropy, $1/\varepsilon \approx 10$.\ Also, we assume pinning by point
defects,\ neglecting the intrinsic pinning in between the Cu-O layers, as
the angular deviation between different experiments in our setup is larger
than the lock-in angle $(\vartheta _{L}<1^{o})$\ \cite%
{blatter,feinberg,Zhukov}. Equation (1) was solved numerically, for $%
\varepsilon =16\pi ^{2}\lambda _{o}^{2}k/\Phi _{o}^{5/2}c_{L}^{2}$ by
inserting the explicit temperature dependences of the coherence length $\xi
=\xi _{o}\left( 1-\left( T/T_{c}\right) ^{4}\right) ^{-1/2}$, the
penetration depth $\lambda =\lambda _{o}\left( 1-\left( T/T_{c}\right)
^{4}\right) ^{-1/2}$, and the pinning parameter $\gamma =\gamma _{o}\left(
1-\left( T/T_{c}\right) ^{4}\right) ^{2}$\cite{giller97}. This procedure
yields the temperature dependence of the order-disorder transition line B$%
_{OD}$(T) for different amplitudes of the pinning parameter $\gamma _{o}$,
as illustrated in Fig. 3. The 'pure' melting line in the figure is obtained
by neglecting the pinning energy, so that $E_{el}=kT$ , whereas the 'pure'
solid-solid transition line is obtained by neglecting the thermal energy,
i.e. when $E_{pin}=E_{el}$ .{\rm \ }All lines in between these two represent
order-disorder transition lines in which {\it both} thermal and pinning
energies are taken into account. Thus, by tuning the pinning strength one
may gradually change the shape of the transition line and the nature of the
disordered phase. In particular, when $E_{pin}$ and $kT$ are comparable, the
behavior of the transition line is qualitatively similar to that of a
melting line, however it represents a transition to a disordered state
exhibiting irreversible magnetic behavior. One may refer to this disordered
state as a 'pinned liquid state' \cite{footnote3}. Our experimental results
for B$_{OD}$(T) in LaSCO, see Fig. 1, clearly indicate that our LaSCO sample
provides an example of a transition into a vortex pinned liquid state driven
by both thermally- and disorder-induced fluctuations.

An indication for the nature of this phase transition was obtained from
partial hysteresis loop measurements \cite{radzyner,roy,kokkaliaris}. These
partial loops exhibit history dependent phenomena in the region $%
H_{onset}^{-}(T)<H<H_{kink}^{+}(T)$, similar to those obtained in YBCO \cite%
{radzyner}. The observed history phenomena indicate that a disordered vortex
state can be ''supercooled'' to exist as a metastable state {\it below} the
transition line, i.e. in the region $H_{onset}^{-}(T)<H<H_{kink}^{-}(T)$.
Likewise, the ordered phase can be ''superheated'' to exist as a metastable
state {\it above} the transition line, in the region $%
H_{kink}^{-}(T)<H<H_{kink}^{+}(T)$. These observations indicate the first
order nature \cite{radzyner,roy2000} of the transition to the vortex
pinned-liquid state. The first order nature of both the melting \cite%
{zeldov1}, and the solid-solid transition \cite{radzyner,roy2000,vanderBeek}%
, was noted previously.

In summary, we observe puzzling temperature dependence of the order-disorder
transition field in LaSCO.\ We show that this behavior may be explained
assuming that both\ thermally- and disorder-induced fluctuations act
together in destroying the ordered phase. This approach leads to the
conclusion that the melting, solid-solid, and solid to pinned-liquid vortex
phase transitions are different manifestations of the same order-disorder
thermodynamic first order transition, which, in general, is driven by {\it %
both} thermally- and disorder-induced fluctuations. This conclusion is in
accordance with several recent works in BSCCO \cite{vanderBeek,avraham},
claiming that the vortex melting line and solid-solid transition line are
two manifestations of the same first order transition. Our results show that
the behavior of the transition line and the nature of the disordered state
are determined by the relative contribution of the disorder-induced
fluctuations. When this contribution is negligible (dominates), a transition
to a liquid (solid) disordered state is obtained. When the contributions of
thermally- and disorder-induced fluctuations are comparable, a transition to
a pinned liquid state is obtained. Thus, the observed transition line
retains the shape of the melting transition, but the pinning suffices for
the transition to be observed as a second peak, and not as a jump in
magnetization. The vortex system in LaSCO exhibits such a transition over a
wide region of the phase diagram.

Acknowledgments. This manuscript is part of Y.R.'s PhD thesis. Continuous
and helpful discussions with D. Giller, A. E. Koshelev and Y. Wolfus are
gratefully acknowledged. Important comments from E. Zeldov, V. Geshkenbein,
T. Giamarchi and A. A. Zhukov are acknowledged. Y. R. acknowledges financial
support from Mifal Hapayis - Michael Landau Foundation. Y. Y. acknowledges
support from the US-Israel Binational Science Foundation. A. S. and I. F.
acknowledge support from the Israel Science Foundation. This research was
supported by The Israel Science Foundation - Center of Excellence Program,
and by the Heinrich Hertz Minerva Center for High Temperature
Superconductivity.

Fig. 1: Temperature dependence of $H_{onset}^{+}(T)$ (up triangles), $%
H_{kink}^{+}(T)$\ (circles), $H_{kink}^{-}(T)$\ (solid diamonds), $%
H_{onset}^{-}(T)$ (up triangles) and the irreversibility line (open
squares). Lines are guides to the eye. Inset: Magnetization loops with the
field parallel to the ab planes, at 12, 16, 20 and 24 K. Arrows point to
four characteristic features plotted in the main panel.

Fig. 2: Relaxation measurements at 12 K, on the ascending branch of the
loop. Grey columns represent measurements extended over an hour. Lines
connect magnetization at t=0 and t=1 hr. Arrows point at the location of the
characteristic features. Note that $H_{onset}^{+}$ shifts about 1 kOe, but $%
H_{kink}^{+}$\ is unaffected. Inset: Dependence of the relaxation rate on
field.

Fig. 3: Numerical solution of $E_{el}=E_{pin}+kT$. The melting (solid-solid
transition) line is calculated by neglecting pinning (thermal) energy. All
lines in between represent order-disorder transition lines in which both
thermal and pinning energies are taken into account, but differ in the
pinning strength, $\gamma _{o\text{ }}$(arbitrary units).


\begin{references}
\bibitem{footnote1} Owing to the small value of the anisotropy, $%
1/\varepsilon \approx 10$, the vortices permeating the sample are not
Josephson vortices, therefore the theoretical approach [7-9] may be applied.
\bibitem{footnote2} Consistently with YBCO [5] the relaxation rate decreases
with field at different rates around the order-disorder transition line. At
a field located between the kink and peak $s=d(\ln m)/d(\ln t)$ has a
minimum (not shown), where a crossover in dynamics probably occurs.
\end{references}
\end{document}